# NLTE Analysis of Copper Abundances in the Galactic Bulge Stars

X. D. Xu[1,2], J. R. Shi[1,2], and H. L. Yan[1,2]
[1] Key Laboratory of Optical Astronomy, National Astronomical Observatories, Chinese Academy of Science, Beijing 100012, People's Republic of China
sjr@bao.ac.cn
[2] University of Chinese Academy of Sciences, Beijing 100049, People's Republic of China

## Abstract

Based on the medium-high resolution ($R \sim 20,000$), modest signal-to-noise ratio (S/N $\gtrsim$ 70) FLAMES-GIRAFFE spectra, we investigated the copper abundances of 129 red giant branch stars in the Galactic bulge with [Fe/H] from $-1.14$ to 0.46 dex. The copper abundances are derived from both local thermodynamic equilibrium (LTE) and nonlocal thermodynamic equilibrium (NLTE) with the spectral synthesis method. We find that the NLTE effects for Cu I lines show a clear dependence on metallicity, and they gradually increase with decreasing [Fe/H] for our sample stars. Our results indicate that the NLTE effects of copper are important not only for metal-poor stars but also for supersolar metal-rich ones and the LTE results underestimate the Cu abundances. We note that the [Cu/Fe] trend of the bulge stars is similar to that of the Galactic disk stars spanning the metallicity range of $-1.14 <$ [Fe/H] $< 0.0$ dex, and the [Cu/Fe] ratios increase with increasing metallicity when [Fe/H] is from $\sim -1.2$ to $\sim -0.5$ dex, favoring a secondary (metallicity-dependent) production of Cu.

## 1. Introduction

Being a major component of the Milky Way Galaxy, the origin and subsequent evolutions of the Galactic bulge are essential both for constraining the Galaxy theoretical evolutionary models, and for interpreting observations of extragalactic objects. Theoretically, many chemical elements can be formed in multiple astrophysical sites and on different timescales. Therefore, the observational behaviors of elemental abundance ratios in diverse stellar populations for a given Galactic region (e.g., the Galactic bulge) allow one to trace the chemical enrichment history of that region, and the elemental abundance ratios as a function of metallicity or time can also afford significant proofs of possible nucleosynthesis processes (Tinsley 1979).

Copper is a peculiar iron-peak element that can be synthesized by several nucleosynthesis scenarios (Bisterzo et al. 2004), and the respective importance of the miscellaneous scenarios is still in dispute. The first one is the weak s-process, which operates in massive stars during core-helium and carbon-shell burning stages, as well as in the explosive complete Ne burning stage (Woosley & Weaver 1995; Limongi & Chieffi 2003; Pignatari et al. 2010); the second possibility is the explosive nucleosynthesis in SNe II (Timmes et al. 1995); the third mechanism can be the main s-process occurring in low and intermediate mass stars during the asymptotic giant branch (AGB; Arlandini et al. 1999), and the last source is the explosive nucleosynthesis taking place in long-lived SNe Ia (Matteucci et al. 1993; Iwamoto et al. 1999; Travaglio et al. 2004; Fink et al. 2014).

So far, many analyses of Cu abundances have been presented, which cover a broad range of metallicity, from the extreme metal-poor stars to solar ones. Generally, the Galactic disk/halo trend ([Cu/Fe] ratio with [Fe/H]) has several interesting features. The [Cu/Fe] trend is roughly flat from solar [Fe/H] down to about [Fe/H] = $-0.7$ dex (e.g., Reddy et al. 2003, 2006; Ishigaki et al. 2013), and then declines linearly with a slope close to 1 with decreasing [Fe/H] in the metallicity range of $-0.7 >$ [Fe/H] $> -1.5$ dex (e.g., Gratton & Sneden 1988; Sneden & Crocker 1988; Sneden et al. 1991; Mishenina et al. 2002; Simmerer et al. 2003; Yan et al. 2015, 2016), while a flat plateau around [Cu/Fe] $\approx -0.7$ to $-1.0$ dex exhibits when [Fe/H] $< -1.5$ dex, (e.g., Westin et al. 2000; Cowan et al. 2002; Sneden et al. 2003; Bihain et al. 2004; Lai et al. 2008).

However, all of the works mentioned above were focused on analyzing the Galactic disk or halo stars, and only a few analyses of Cu abundances in bulge stars were carried out. Johnson et al. (2014) derived the Cu abundances of a large sample of bulge red giant branch stars and found that the bulge had a distinct [Cu/Fe] trend with [Fe/H], very different from that of the Galactic disk and halo stars. In the bulge, the Cu abundances increased monotonically from [Cu/Fe] = $-0.84$ dex in the most metal-poor stars to [Cu/Fe] $\sim +0.4$ dex in the most metal-rich ones. Ernandes et al. (2018) analyzed the Cu abundances in individual stars (metallicity of a range of $-1.2 \leqslant$ [Fe/H] $\leqslant 0.0$) of five bulge globular clusters (HP 1, NGC 6522, NGC 6528, NGC 6553, and NGC 6558), and noted that the abundances of Cu showed a behavior of secondary element, and were in good agreement with predictions of the chemical evolution models by Kobayashi et al. (2006). Reviews on copper abundances in the Galactic bulge can be found in McWilliam (2016) and Barbuy et al. (2018).

For Johnson et al.'s (2014) results, McWilliam (2016) raised an important question about whether the large scatter in the [Cu/Fe] of metal-rich bulge stars was real, and suggested that the nonlocal thermodynamic equilibrium (NLTE) corrections for Cu should be considered for these objects. NLTE effects, generally neglected, can affect Cu abundance calculations as a function of surface temperature, gravity, and/or metallicity of stars. Recent studies on NLTE effects had demonstrated that the NLTE corrections were large for Cu, particularly in metal-poor stars (e.g., Shi et al. 2014, 2018; Yan et al. 2015, 2016; Andrievsky et al. 2018). Clearly, more work is needed to better investigate the Cu abundances under NLTE assumption for bulge stars, especially for the metal-rich ones.

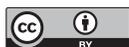

In this paper, we present the local thermodynamic equilibrium (LTE) and NLTE copper abundances for a sample of stars in the bulge, by applying a complete spectrum synthesis method based on level populations calculated from the statistical equilibrium equations. In Section 2, the observations are briefly summarized. In Section 3, the adapted model atmospheres, stellar parameters, and atomic line data are reported. Then, the copper atomic model and the NLTE effects are described in Section 4. In Section 5, the results and discussion are presented, including the derived copper abundances, the comparison with other works, and the evolutionary trend of [Cu/Fe] in the bulge. Finally, the conclusions are provided in Sections 6.

## 2. Observations

Our sample consists of 129 red giant branch stars in the Galactic bulge, and they have already been minutely discussed by Zoccali et al. (2008) and Johnson et al. (2014). Here, we briefly summarize the key points of the observations. Details regarding the selection of targets and input parameters are given in the aforementioned papers.

1. The observations were carried out at the VLT-UT2 with the FLAMES-GIRAFFE spectrograph, at a resolution power $\sim$20,000. The original program by Zoccali et al. (2008) contained four fields, but as the HR11 (spanning $\sim$5590–5835 Å) was the only setup including measurable copper lines, only two bulge fields, (+5.25, −3.02) and (0, −12), were observed in the HR11, HR13, and HR15 setups (ESO[3] Program 073.B-0074).
2. Johnson et al. (2014) analyzed 156 giants with the final signal-to-noise ratio (S/N) of each coadded spectrum exceeding $\sim$70, and without strong TiO absorption bands. As the spectral S/Ns were rather low for some stars, the final sample utilized here only included 129 objects.
3. The raw data had been reduced with the GIRAFFE pipeline,[4] including bias subtraction, flat-field correction, extraction, and wavelength-calibration, by the Quality Control Group of ESO, and the reduced spectra were directly downloaded from the ESO Science Archive Facility. All the spectra for each star (a number between one and five) were then radial velocity corrected and coadded to a combined spectrum, and continuum normalized via the SIU code (Reetz 1991).

## 3. Method of Calculation

### 3.1. Model Atmospheres

An appropriate stellar atmospheric model is the basis of the spectrum synthesis method, and is important for high-quality abundance determination. In this study, we adopted the MARCS atmospheric models, and they were obtained by interpolation in the grid of spherical or plane-parallel MARCS models ($-1.0 \leqslant \log g \leqslant 3.5$ for spherical models; $3.0 \leqslant \log$

---

[3] European Southern Observatory.
[4] http://www.eso.org/sci/software/pipelines/

**Table 1**
Atomic Data of the Copper Line

| $\lambda_{\mathrm{air}}$ (Å) | Transition | $E_{\mathrm{low}}$ (eV) | $\log gf$ | $\log C_6$ |
|---|---|---|---|---|
| 5782.132 | $4s^2\ ^2D_{3/2}$–$4p\ ^2P^o_{1/2}$ | 1.642 | −1.89 | −31.66 |

**Note.** The log $gf$ value was rectified by fitting the NLTE solar spectrum. The van der Waals damping constant (log $C_6$) was calculated according to Anstee & O'Mara (1991, 1995).

$g \leqslant 5.5$ for plane-parallel models; see Gustafsson et al. 2008 for details). For each individual model, the enhanced $\alpha$-elements were used. As assumed by Fulbright et al. (2007), [Mg/Fe] was an appropriate surrogate for the [$\alpha$/Fe] ratio, which was better than that from a mean of several alpha-element ratios. Therefore, these models considered [Mg/Fe], derived by Johnson et al. (2014), as the [$\alpha$/Fe].

### 3.2. Stellar Parameters

We directly employed the stellar parameters given in Johnson et al. (2014) for all of our program stars. A brief description of the method is as follows:

1. The stellar parameters were obtained via spectroscopic analysis, and the procedure of determining stellar parameters was an iterative process. Four primary input parameters (effective temperature ($T_{\mathrm{eff}}$), surface gravity (log $g$), metallicity ([Fe/H]), and micro-turbulence ($\xi$)) were needed to begin with before converging to a solution. The model parameters given in Zoccali et al. (2008) were used for stars in the (0, −12) field, and generic values of $T_{\mathrm{eff}}$ = 4500 K, log $g$ = 2.0 dex, [Fe/H] = −0.20 dex, and $\xi$ = 1.5 km s$^{-1}$ were adopted for stars in the (+5.25, −3.02) field.
2. The final $T_{\mathrm{eff}}$ and log $g$ were determined by enforcing the excitation equilibrium of Fe I and the ionization equilibrium between Fe I and Fe II, respectively; the $\xi$ was determined by removing trends in Fe I abundance versus their equivalent widths.

### 3.3. Atomic Line Data

We used the Sun as a reference star for the subsequent abundance analysis, and adopted the absolute solar copper abundance derived by Lodders et al. (2009), which was log $\varepsilon_\odot$(Cu) = 4.25. Since the copper line at 5700 Å was seriously blended with other lines and its profile was irregular, only one measurable copper line at 5782 Å was used in this analysis. Table 1 presents the line data with its final solar fitted $gf$ value (Shi et al. 2014). In this table, the log $gf$ value had been rectified from the NLTE solar spectrum fitting. The van der Waals damping constant (log $C_6$) was computed with Anstee & O'Mara (1991, 1995) tables. We assumed that the ratio between two copper isotopes ($^{63}$Cu and $^{65}$Cu) was 0.69:0.31 (Asplund et al. 2009). The hyperfine structure, calculated according to Biehl (1976) data, was also taken into account in our work.

## 4. NLTE Calculations

### 4.1. Atomic Model

For our NLTE calculations, we adopted the copper atomic model from Shi et al. (2014). The model contained 97 energy levels (96 states for Cu I and the ground state of Cu II), and 1089 radiative and collisional transitions. Inelastic collisions with electrons were calculated according to several theoretical works (Seaton 1962; van Regemorter 1962; Allen 1973). The inelastic collisions with hydrogen atoms were obtained based on the Drawins formula (Drawin 1968, 1969) used by Steenbock & Holweger (1984) with a hydrogen collision enhancement factor ($S_H = 0.1$) under the suggestion of Shi et al. (2014). In addition, we introduced the photoionization cross sections and the oscillator strengths given by Liu et al. (2014) to our atomic model.

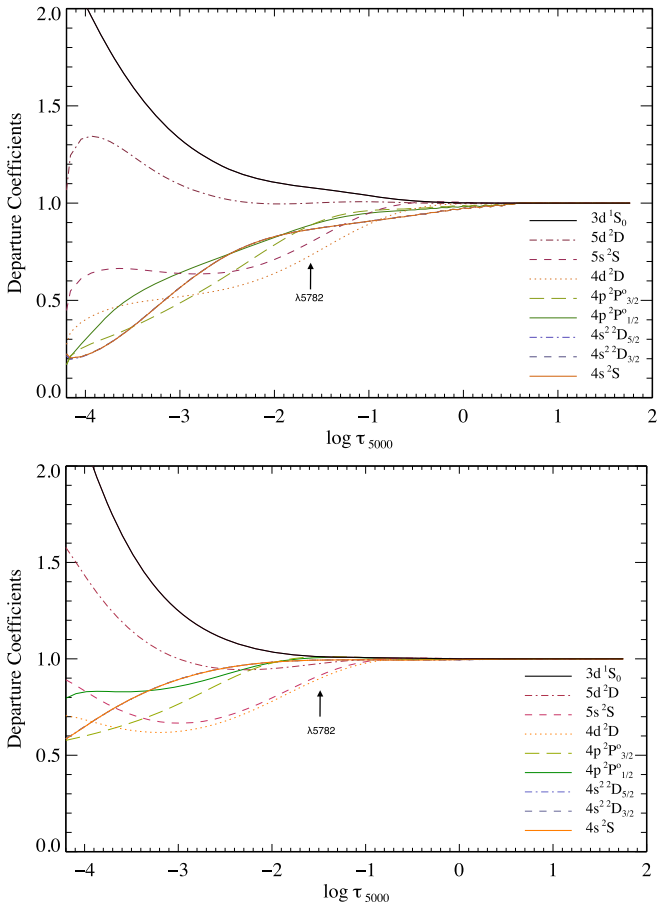

**Figure 1.** Departure coefficients ($b_i$) for important Cu I energy levels and the Cu II ground state as a function of continuum optical depth at 5000 Å for the model atmospheres of 1918C1 (top panel) and 6230C5 (bottom panel).

As NLTE calculations required solving the coupled radiative transfer and statistical equilibrium equations, we adopted a revised DETAIL program (Butler & Giddings 1985) based on an accelerated lambda iteration method (Rybicki & Hummer 1991, 1992) to compute the NLTE deviations in the level populations for the Cu I model. Then, the obtained departure coefficients were used to calculate the synthetic spectra with the SIU code (Reetz 1991).

### 4.2. NLTE Effects

In Figure 1, we plot the departure coefficients ($b_i$) versus the optical depth at $\lambda = 5000$ Å (log $\tau_{5000}$) for the model atmospheres of 1918C1 and 6230C5. As usual, we define the departure coefficients ($b_i$) as $b_i = (n_i^{\mathrm{NLTE}}/n_i^{\mathrm{LTE}})$, where $n_i^{\mathrm{NLTE}}$ and $n_i^{\mathrm{LTE}}$ represent the statistical and thermal equilibrium atomic level populations, respectively. The departure coefficients for some important Cu I levels and the Cu II ground state are presented in this figure. For the convenience of discussion, we take two stars, 1918C1 and 6230C5, as examples. 1918C1 is a typical metal-poor star with moderate temperature among our sample, while 6230C5 is a supersolar metallicity star with a higher temperature. Figure 1 clearly shows that the Cu II 3$d$ $^1S_0$ ground state is overpopulated in the deep atmosphere due to overionization for both stars, and the two low excited levels 4$s^2$ $^2D_{5/2}$ and 4$s^2$ $^2D_{3/2}$ overlap with the Cu I ground state 4$s$ $^2S$. However, from the b-factor's behavior of 4$s$ $^2S$ level, we can still see that the number densities of the lower (4$s^2$ $^2D_{3/2}$) and upper (4$p$ $^2P^o_{1/2}$) levels for the 5782 Å line begin to be underpopulated owing to overionization beyond log $\tau_{5000} \sim 0.5$. In the range of log $\tau_{5000} < -3$, an obvious overpopulation of the level 5$d$ $^2D$, which results from over-recombination, can also be seen.

In order to investigate the correlation between NLTE effects and stellar parameters, we display in Figure 2 how the NLTE corrections, $\Delta_{\mathrm{NLTE}} = [\mathrm{Cu/Fe}]_{\mathrm{NLTE}} - [\mathrm{Cu/Fe}]_{\mathrm{LTE}}$, vary with effective temperature, surface gravity, and metallicity, respectively. The results are best characterized by a clear metallicity dependence on the NLTE effects and the $\Delta_{\mathrm{NLTE}}$ gradually increases with decreasing [Fe/H]. The $\Delta_{\mathrm{NLTE}}$ is positive, which signifies that the LTE assumption will underestimate the copper abundances. We note that the NLTE effects may be slightly sensitive to the surface gravity with a tendency of increasing with decreasing log $g$, while there is no clear trend in the NLTE effects with effective temperature. Theoretically, if the [Fe/H] and log $g$ decrease, UV line blocking will decrease rapidly and the overionization will become more important. Besides, for late-type giants, the densities of electron and neutral hydrogen are low and the electronic collisions become less important. Therefore, the NLTE effects, as expected, will increase with decreasing [Fe/H] and/or log $g$.

### 5. Results and Discussion

#### 5.1. Stellar Copper Abundances

Throughout this study, the copper abundances of our program stars were derived using the spectral synthesis method under both LTE and NLTE assumptions. Abundance determination by this method was an iterative process. We repeatedly modified the copper abundance in each step of the process until the synthetic spectrum reached the best fitting compared to the observed one. In order to fit the observed spectral lines, we treated the broadenings caused by the rotation, macroturbulence, and instrument as one single Gauss profile to be convolved with the synthetic spectra.

Figure 3 shows the comparison between the observed and synthetic line profiles for 1918C1 and 6230C5. It is obvious that the NLTE calculations agree well with the observed spectrum in the line core as well as in the wing, while the LTE line profiles are substantially deeper in the line core. In Table 2, we present the final derived LTE and NLTE copper abundances. It can be seen that the abundances derived from NLTE are larger than those from LTE, and the NLTE corrections are evident with a variation from 0.0 to ~0.2 dex. The uncertainty in the fit is about 0.1 dex in the derived copper abundance, and the uncertainty caused by random errors on the determination of the stellar parameters for a mildly metal-deficient star 1918C1 is given in Table 3.

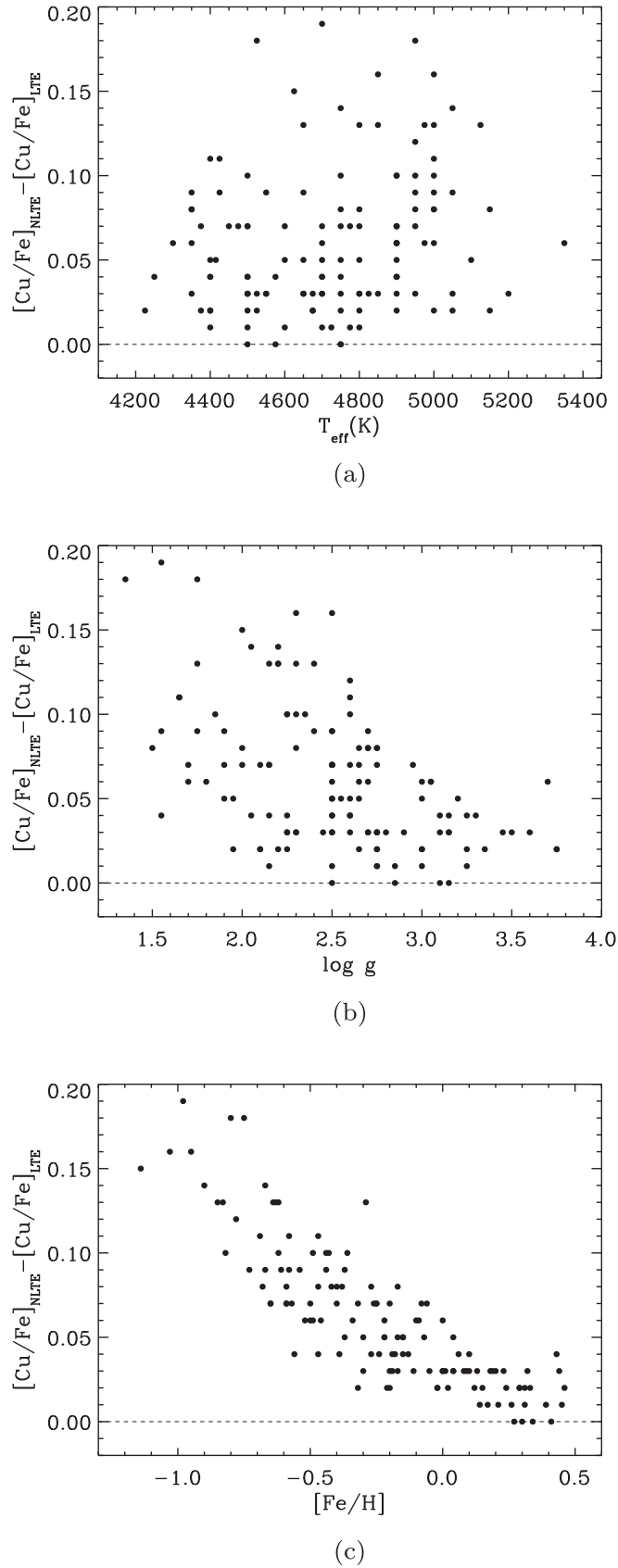

**Figure 2.** Differences of [Cu/Fe] abundance ratios between LTE and NLTE as a function of effective temperature (a), surface gravity (b), and metallicity (c).

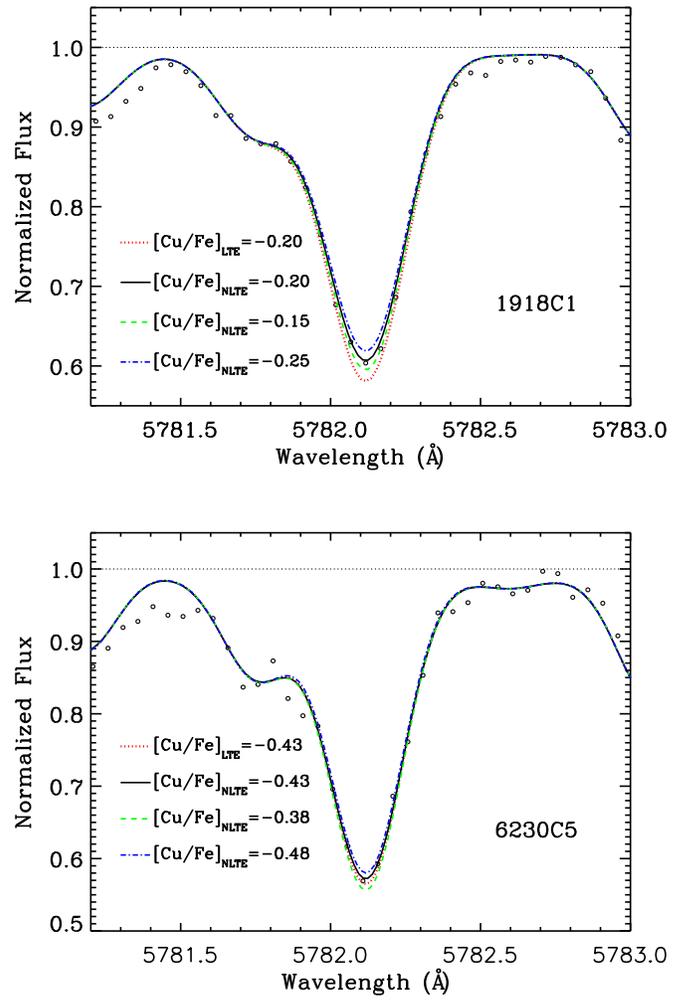

**Figure 3.** Synthetic profiles of Cu I line at 5782 Å for 1918C1 and 6230C5. The open circles are the observational data, the black solid line is the best-fit synthesis (NLTE synthesis), the red dotted line is the LTE synthesis with the same [Cu/Fe] relative to NLTE, and the green dashed and blue dashed–dotted lines are synthetic spectra with [Cu/Fe] of ±0.05 dex relative to the best fit.

### 5.2. Comparison with Other Work

To date, only two groups have determined copper abundances for the bulge stars based on LTE analysis. Johnson et al. (2014) derived copper abundances for a large sample of bulge stars, and they found that the [Cu/Fe] ratios of bulge stars increased with higher [Fe/H]. At the supersolar metallicity region, the [Cu/Fe] ratios appeared to be much enhanced, relative to the Galactic disk stars. A recent study by Ernandes et al. (2018) analyzed the Cu abundances in individual stars of five bulge globular clusters; however, their results showed that the [Cu/Fe] ratios were less enhanced than those by Johnson et al. (2014).

We compare our LTE results to those from Johnson et al. (2014), and display the differences as a function of metallicity for the same objects in Figure 4. It is surprising that the

Table 2
Copper Abundances of Our Program Stars

| Star | $T_{\text{eff}}$ | log g | [Fe/H] | $\xi$ | [Cu/Fe]$_{\text{LTE}}$ | [Cu/Fe]$_{\text{NLTE}}$ |
|---|---|---|---|---|---|---|
| (l, b) = (+5.25, −3.02) | | | | | | |
| 119799C4 | 4400 | 2.05 | −0.39 | 1.90 | −0.08 | −0.04 |
| 129499C4 | 4900 | 2.75 | +0.13 | 2.10 | −0.03 | 0.00 |
| 181349C5 | 4775 | 2.75 | +0.26 | 2.10 | −0.05 | −0.04 |
| 184088C5 | 4500 | 2.50 | −0.47 | 1.80 | 0.14 | 0.18 |
| 184618C5 | 4600 | 2.15 | −0.59 | 1.55 | −0.10 | −0.03 |
| 185169C5 | 4775 | 2.50 | −0.32 | 1.85 | 0.00 | 0.07 |
| 185541C5 | 4750 | 3.10 | +0.34 | 1.80 | 0.05 | 0.05 |
| 187067C5 | 4850 | 2.30 | −0.95 | 1.35 | −0.31 | −0.15 |
| 193190C5 | 4675 | 2.75 | +0.24 | 1.85 | −0.06 | −0.04 |
| 197366C5 | 4550 | 2.90 | +0.32 | 1.75 | 0.25 | 0.28 |
| 215681C6 | 4800 | 2.75 | +0.31 | 1.95 | −0.13 | −0.12 |
| 216922C6 | 4400 | 2.10 | −0.20 | 2.20 | −0.05 | −0.03 |
| 218198C6 | 4225 | 1.95 | −0.32 | 2.10 | −0.06 | −0.04 |
| 221537C6 | 4750 | 2.75 | +0.29 | 1.60 | 0.07 | 0.09 |
| 223113C6 | 4800 | 2.55 | −0.15 | 1.75 | 0.08 | 0.13 |
| 223343C6 | 4525 | 2.65 | +0.15 | 1.45 | −0.20 | −0.18 |
| 223621C6 | 5150 | 3.75 | +0.31 | 1.35 | 0.21 | 0.23 |
| 223722C6 | 4525 | 1.35 | −0.75 | 1.75 | 0.02 | 0.20 |
| 224206C6 | 4800 | 3.00 | +0.46 | 1.65 | 0.15 | 0.17 |
| 225531C6 | 4425 | 1.75 | −0.73 | 1.80 | −0.10 | −0.01 |
| 226450C6 | 4500 | 2.50 | +0.27 | 1.80 | −0.05 | −0.05 |
| 226850C6 | 4800 | 2.15 | −0.85 | 1.85 | −0.10 | 0.03 |
| 227867C6 | 4850 | 3.15 | +0.09 | 1.60 | 0.07 | 0.10 |
| 228466C6 | 4700 | 1.55 | −0.98 | 2.15 | −0.05 | 0.14 |
| 229507C6 | 4650 | 1.75 | −0.29 | 1.60 | −0.01 | 0.12 |
| 230483C6 | 4450 | 2.10 | −0.65 | 1.60 | 0.06 | 0.13 |
| 231379C6 | 4500 | 2.30 | −0.30 | 1.80 | −0.04 | −0.01 |
| 231618C6 | 4550 | 2.60 | +0.04 | 1.65 | −0.16 | −0.13 |
| 232493C6 | 4625 | 2.00 | −1.14 | 1.20 | −0.50 | −0.35 |
| 233121C6 | 4500 | 1.90 | −0.50 | 1.90 | 0.07 | 0.14 |
| 233560C6 | 4375 | 2.10 | −0.21 | 1.85 | −0.18 | −0.16 |
| 233708C6 | 4500 | 2.50 | +0.14 | 1.70 | −0.25 | −0.24 |
| 240059C6 | 4400 | 1.65 | −0.69 | 1.60 | 0.09 | 0.20 |
| 240083C6 | 4425 | 1.65 | −0.58 | 1.70 | 0.09 | 0.20 |
| 259377C7 | 5000 | 3.25 | +0.29 | 1.85 | 0.12 | 0.14 |
| 262018C7 | 4700 | 2.75 | +0.08 | 2.00 | −0.03 | 0.00 |
| 266442C7 | 4350 | 2.00 | −0.42 | 1.60 | 0.18 | 0.26 |
| 270316C7 | 4400 | 1.90 | −0.37 | 1.65 | −0.14 | −0.09 |
| 275181C7 | 4375 | 1.70 | −0.59 | 1.75 | −0.16 | −0.09 |
| 277490C7 | 4250 | 1.55 | −0.56 | 1.65 | −0.29 | −0.25 |
| 282804C7 | 4575 | 2.50 | +0.10 | 1.70 | −0.17 | −0.13 |
| 286252C7 | 4700 | 3.10 | +0.43 | 1.40 | 0.34 | 0.38 |
| 45512C2 | 4500 | 2.15 | −0.27 | 1.75 | −0.14 | −0.10 |
| 47188C2 | 4350 | 1.70 | −0.52 | 1.85 | −0.03 | 0.03 |
| 77186C3 | 4750 | 3.15 | +0.41 | 1.85 | 0.04 | 0.04 |
| 77707C3 | 4825 | 3.15 | +0.44 | 1.65 | 0.08 | 0.11 |
| 81644C3 | 4600 | 2.85 | +0.21 | 1.70 | −0.12 | −0.11 |
| 83531C3 | 4725 | 3.25 | +0.45 | 1.80 | 0.11 | 0.12 |
| 84255C3 | 4350 | 1.50 | −0.68 | 1.45 | −0.28 | −0.20 |
| 86757C3 | 4400 | 2.20 | −0.02 | 1.90 | −0.26 | −0.24 |
| 88522C3 | 4700 | 3.00 | +0.39 | 1.90 | 0.00 | 0.01 |
| 225847C6 | 4500 | 2.25 | −0.11 | 1.45 | −0.20 | −0.17 |
| 227379C6 | 4350 | 2.25 | −0.05 | 1.55 | −0.22 | −0.19 |
| 228407C6 | 4415 | 1.95 | −0.22 | 1.75 | 0.02 | 0.07 |
| 230208C6 | 4900 | 3.05 | +0.00 | 1.40 | 0.17 | 0.23 |
| 265795C7 | 4600 | 2.50 | −0.07 | 1.55 | 0.18 | 0.23 |
| 268493C7 | 4500 | 2.30 | −0.20 | 1.90 | 0.07 | 0.10 |
| 271021C7 | 4750 | 2.50 | −0.06 | 1.65 | 0.11 | 0.18 |
| 271400C7 | 4525 | 2.50 | −0.19 | 1.90 | −0.07 | −0.04 |
| 77182C3 | 4700 | 2.50 | −0.09 | 1.75 | 0.18 | 0.24 |
| 85597C3 | 4400 | 2.25 | −0.15 | 1.55 | −0.02 | 0.02 |
| (l, b) = (0, −12) | | | | | | |

Table 2
(Continued)

| Star | $T_{\rm eff}$ | log g | [Fe/H] | $\xi$ | [Cu/Fe]$_{\rm LTE}$ | [Cu/Fe]$_{\rm NLTE}$ |
|---|---|---|---|---|---|---|
| 1156C2 | 4300 | 1.80 | −0.46 | 1.45 | −0.16 | −0.10 |
| 1407C3 | 5125 | 2.40 | −0.62 | 1.45 | −0.24 | −0.11 |
| 1554C7 | 5050 | 2.20 | −0.67 | 1.50 | −0.09 | 0.05 |
| 166C3 | 4850 | 2.20 | −0.83 | 1.30 | −0.46 | −0.33 |
| 1754C3 | 4900 | 2.50 | −0.25 | 1.65 | −0.12 | −0.05 |
| 1814C1 | 4650 | 2.60 | −0.30 | 1.50 | −0.10 | −0.05 |
| 1876C2 | 5000 | 2.50 | −1.03 | 1.10 | −0.42 | −0.26 |
| 1917C1 | 4675 | 2.70 | +0.18 | 1.50 | −0.22 | −0.19 |
| 1918C1 | 4900 | 2.35 | −0.44 | 1.60 | −0.05 | 0.05 |
| 2110C7 | 4675 | 3.00 | +0.02 | 1.30 | −0.07 | −0.05 |
| 2178C7 | 4700 | 2.80 | +0.00 | 1.80 | −0.20 | −0.17 |
| 2200C3 | 4975 | 2.70 | −0.10 | 1.80 | −0.15 | −0.09 |
| 2220C7 | 5150 | 2.75 | −0.17 | 1.60 | 0.09 | 0.17 |
| 222C3 | 4950 | 2.60 | −0.43 | 1.90 | 0.09 | 0.19 |
| 2335C2 | 4750 | 2.05 | −0.90 | 1.50 | −0.13 | 0.01 |
| 2407C2 | 4975 | 2.30 | −0.64 | 1.70 | −0.10 | 0.03 |
| 2422C7 | 4400 | 2.15 | +0.17 | 1.90 | −0.31 | −0.30 |
| 2470C3 | 4800 | 3.10 | +0.00 | 1.35 | 0.02 | 0.05 |
| 2502C3 | 4750 | 2.25 | −0.62 | 1.40 | −0.22 | −0.12 |
| 2532C6 | 4750 | 2.50 | −0.17 | 1.90 | −0.20 | −0.15 |
| 2580C6 | 4575 | 2.85 | +0.30 | 2.00 | −0.26 | −0.26 |
| 2769C3 | 4900 | 3.15 | +0.06 | 1.25 | −0.10 | −0.06 |
| 2772C7 | 4900 | 2.75 | −0.25 | 1.65 | 0.11 | 0.18 |
| 2812C8 | 4475 | 2.00 | −0.65 | 1.55 | −0.32 | −0.25 |
| 3018C3 | 4650 | 2.40 | −0.54 | 1.45 | 0.02 | 0.11 |
| 3035C7 | 5050 | 2.50 | −0.44 | 1.45 | −0.19 | −0.10 |
| 3091C8 | 4900 | 3.00 | −0.50 | 1.45 | −0.04 | 0.02 |
| 3101C7 | 4950 | 2.65 | −0.40 | 1.70 | −0.07 | 0.01 |
| 3142C3 | 4900 | 2.65 | −0.22 | 1.75 | −0.01 | 0.05 |
| 3161C3 | 5100 | 3.20 | −0.15 | 1.70 | 0.17 | 0.22 |
| 3191C7 | 4950 | 2.50 | −0.37 | 1.60 | 0.00 | 0.09 |
| 3201C6 | 5200 | 3.50 | +0.04 | 1.35 | −0.08 | −0.05 |
| 3238C6 | 4900 | 3.00 | −0.22 | 1.35 | −0.07 | −0.02 |
| 3267C3 | 4700 | 2.65 | +0.04 | 1.40 | 0.06 | 0.11 |
| 3515C3 | 4750 | 2.70 | +0.01 | 1.80 | −0.15 | −0.12 |
| 3711C7 | 5350 | 3.70 | −0.49 | 1.25 | 0.05 | 0.11 |
| 3733C3 | 4750 | 2.60 | −0.18 | 1.70 | −0.15 | −0.11 |
| 3796C6 | 4500 | 1.85 | −0.82 | 1.65 | −0.19 | −0.09 |
| 4085C3 | 4550 | 1.90 | −0.58 | 1.65 | −0.23 | −0.14 |
| 4217C6 | 4500 | 2.15 | −0.57 | 1.70 | −0.28 | −0.21 |
| 4263C6 | 5000 | 3.05 | −0.34 | 1.90 | 0.15 | 0.21 |
| 4365C3 | 4950 | 1.75 | −0.80 | 1.65 | −0.34 | −0.16 |
| 4478C8 | 4900 | 2.25 | −0.36 | 1.65 | −0.04 | 0.06 |
| 455C1 | 5000 | 2.70 | −0.61 | 1.50 | −0.07 | 0.02 |
| 4612C6 | 4950 | 2.60 | −0.78 | 1.30 | −0.15 | −0.03 |
| 4740C8 | 5000 | 2.30 | −0.49 | 1.60 | −0.16 | −0.06 |
| 4876C6 | 4800 | 2.70 | −0.59 | 1.35 | −0.20 | −0.12 |
| 5351C8 | 4700 | 2.65 | −0.08 | 1.55 | 0.12 | 0.19 |
| 5400C8 | 5050 | 3.75 | +0.12 | 1.30 | −0.04 | −0.02 |
| 5487C8 | 4750 | 2.30 | −0.47 | 1.75 | −0.20 | −0.12 |
| 5588C6 | 4900 | 2.60 | −0.20 | 1.60 | −0.13 | −0.06 |
| 5664C6 | 4800 | 2.75 | +0.23 | 1.60 | −0.21 | −0.18 |
| 5977C6 | 4650 | 2.45 | +0.19 | 1.65 | −0.11 | −0.08 |
| 5980C6 | 4950 | 3.45 | −0.17 | 1.25 | −0.03 | 0.00 |
| 6090C6 | 4500 | 2.25 | −0.02 | 1.90 | −0.22 | −0.20 |
| 6164C6 | 4900 | 3.25 | −0.13 | 0.95 | 0.07 | 0.11 |
| 6230C5 | 5050 | 3.60 | +0.20 | 1.25 | −0.21 | −0.18 |
| 6391C8 | 5000 | 2.20 | −0.63 | 1.55 | −0.01 | 0.12 |
| 6419C5 | 4700 | 2.60 | −0.24 | 1.55 | −0.13 | −0.09 |
| 6426C8 | 4950 | 2.95 | −0.40 | 1.60 | 0.07 | 0.14 |
| 6505C6 | 5000 | 2.70 | −0.38 | 1.50 | 0.09 | 0.17 |
| 650C2 | 4350 | 1.55 | −0.67 | 1.60 | −0.15 | −0.06 |
| 6549C6 | 4900 | 3.35 | +0.33 | 1.30 | 0.00 | 0.02 |
| 6637C8 | 4800 | 2.50 | −0.26 | 1.60 | −0.07 | 0.00 |

**Table 2**
(Continued)

| Star | $T_{\rm eff}$ | log g | [Fe/H] | $\xi$ | [Cu/Fe]$_{\rm LTE}$ | [Cu/Fe]$_{\rm NLTE}$ |
|---|---|---|---|---|---|---|
| 6717C6 | 4900 | 3.30 | −0.19 | 1.20 | 0.04 | 0.08 |
| 6828C7 | 4650 | 2.50 | +0.10 | 1.70 | −0.22 | −0.19 |
| 6913C7 | 5000 | 2.75 | −0.27 | 1.70 | 0.12 | 0.20 |
| 867C3 | 5000 | 2.60 | −0.47 | 1.60 | 0.08 | 0.19 |

**Note.** The information of the stellar parameters are described in the text. The LTE and NLTE results of each star are shown in the last two columns, respectively.

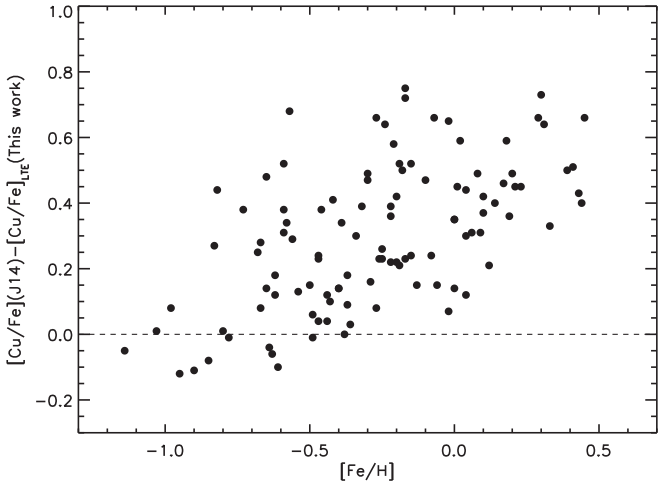

**Figure 4.** Differences of [Cu/Fe] abundance ratios between our work and Johnson et al. (2014) for the same objects as a function of metallicity.

**Table 3**
[Cu/Fe] Uncertainties Linked to Stellar Parameters

|  | $\Delta T$ 100 K | $\Delta$log g 0.3 dex | $\Delta$[Fe/H] 0.15 dex | $\Delta\xi$ 0.3 km s$^{-1}$ |
|---|---|---|---|---|
| $\Delta$[Cu/Fe] | 0.11 | 0.05 | 0.14 | 0.06 |

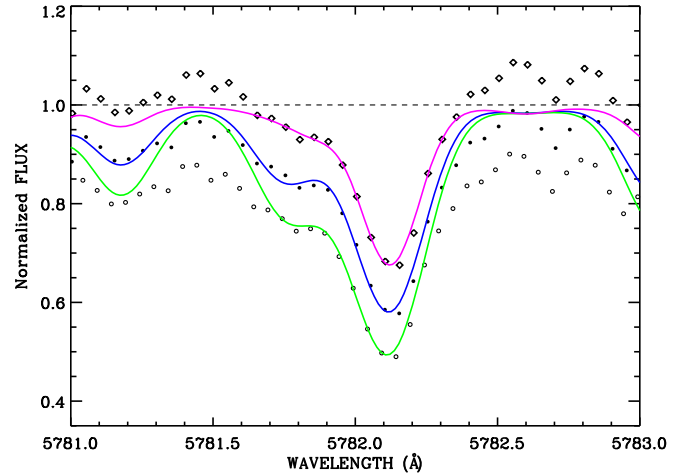

**Figure 5.** Synthetic profiles of the Cu I 5782 Å line for 5980C6. The filled circles are the observational data of the true continuum flux level; the open diamonds and circles are the observational data with about a 10% shift up and down from the true continuum flux level, respectively. The blue, pink, and green solid lines are the best-fit NLTE synthesis.

differences have a linear relationship with [Fe/H], and they tend to be larger for high metallicity stars with the maximum difference of 0.75 dex. As described in Section 5.1, the NLTE corrections vary from 0.0 to ∼0.2 dex. Although our work has taken NLTE effects into account, the large discrepancy between two works still exists, and it cannot just be explained by the NLTE effects.

Although the exact reason for the discrepancy is unclear, we speculate that it could be mainly attributed to the different continuum determination. For all of our program stars, the determination of continuum requires great care because of its large impact on final copper abundances. To present the impact intuitively, we randomly select a typical metal-poor star, 5980C6, for discussion and show its synthetic profiles in Figure 5. It is clear that the impact is really large for determination of [Cu/Fe] ratios. In practice, a 10% shift from the true continuum flux level will yield a systematic change of about 0.3 to 0.6 dex in the [Cu/Fe] values. If the continuum placement is set higher than its true position, copper abundances will be consequently enhanced.

Another possible source for this discrepancy is the negligence of other lines blending. In fact, the Cu I λ5782 line is blended with several other lines (i.e., Cr I, Cr II, Fe I, and Fe II). Among them, most lines show little effect on abundance determination, except for the Cr I line, which blends into the left side of the Cu I line wing. Different abundance values of the blending Cr I line translate into big changes in the Cu abundance of the 5782 Å line, particularly for metal-rich stars. When taking Cr abundance into account and by increasing the [Cr/Fe] value by 0.1 dex, the Cu abundance can be reduced by at least 0.08 dex. In addition, as the Cu I λ5782 line is rather strong in most of our stars, the difference in the adopted van der Waals damping constant (log $C_6$) may also partly contribute to the discrepancy. To explore its influence on our abundance measurements, we randomly select about 20 objects, whose equivalent widths are more than 120 mÅ, for testing. By varying the value of log $C_6$ by 0.3 dex, which is a big change according to our past experience, the derived copper abundances just have negligible changes. Therefore, this cause can definitely be excluded.

### 5.3. The Evolutionary Trend of [Cu/Fe] in the Bulge

The behavior of [Cu/Fe] ratios with the stellar metallicity [Fe/H] holds important information about the sources of copper and the chemical evolution of our Galaxy. Copper can be produced through various nucleosynthetic processes, and it is suggested that a large fraction of Cu is formed in a weak s-process during core He burning, and convective shell C burning phases in massive stars (e.g., Limongi & Chieffi 2003), while the other relevant sources may be SNIa, AGB stars, and SNII (e.g., Timmes et al. 1995; Arlandini et al. 1999; Iwamoto et al. 1999).

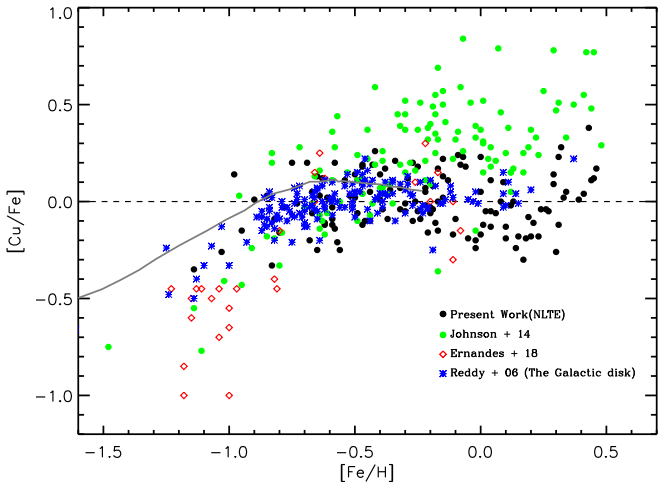

**Figure 6.** [Cu/Fe] vs. [Fe/H] for the present sample (black filled circles), Johnson et al. (2014; green filled circles), Ernandes et al. (2018; red open diamonds) and Reddy et al. (2006; blue asterisks). The theoretical predictions of Kobayashi et al. (2006; gray solid line) are also presented.

We present the [Cu/Fe] ratios versus [Fe/H] for our NLTE calculations (black filled circles), together with LTE results from Reddy et al. (2006) of the Galactic disk (blue asterisks), Johnson et al. (2014; green filled circles) and Ernandes et al. (2018; red open diamonds) of the bulges in Figure 6. It is clear that our calculations tend to be less enhanced than those by Johnson et al. (2014) and exhibit a discrepant [Cu/Fe] trend. We note that there are no obvious differences in copper abundances between our sample and those from Reddy et al. (2006). We also find, in good agreement with Ernandes et al. (2018), that the bulge shows a similar [Cu/Fe] trend with that of the Galactic disk.

The behaviors of our bulge [Cu/Fe] abundance ratios show several features in Figure 6; they increase with increasing metallicity up to [Fe/H] $\sim -0.5$ dex, and then decline monotonically in the metallicity range of $-0.5 <$ [Fe/H] $< +0.2$ dex, similar to the slope of the thin and thick disks, while they again rise with increasing metallicity when [Fe/H] $> +0.2$ dex.

The rise of [Cu/Fe] ratios with increasing metallicity, beyond [Fe/H] $\sim -1$ dex, is most likely due to the metallicity-dependent Cu yield from the weak $s$-process in massive stars (SNII progenitors, e.g., Bisterzo et al. 2004), and the subsequent decline can be the result of the time-delay contribution of Fe from SN Ia (Tinsley 1979; Matteucci & Brocato 1990), while the final rise could be explained by the metal-dependent Cu yields continuing to increase into the super metal-rich regime, and these higher copper yields from SNII progenitors eventually overwhelming the SN Ia iron (McWilliam 2016). Consequently, the trend again increases with increasing [Fe/H].

Many works (e.g., Sneden et al. 1991; Timmes et al. 1995; Goswami & Prantzos 2000; Mishenina et al. 2002; Kobayashi et al. 2006, 2011; Romano & Matteucci 2007; Romano et al. 2010) have attempted to model the Galactic evolution of copper, and all of them are devoted to constructing an accurate Galactic chemical evolution (GCE) model to fulfill the observed data. In Figure 6, we overplot the behavior of copper, predicted by the GCE model from Kobayashi et al. (2006). It is noted that our [Cu/Fe] ratios are well consistent with the model predictions and increase with increasing [Fe/H] throughout the range $\sim -1.2 <$ [Fe/H] $< \sim -0.5$ dex, which suggests that copper seems to behave as a secondary (metallicity-dependent) element. However, Korotin et al. (2018) gave an opposite point of view. Meanwhile, a truncation of the model is also arresting, which may be due to the paucity of [Cu/Fe] ratios at [Fe/H] $> 0$. Since the sample in this work covers more metal-rich stars, our calculations will be helpful to constrain and improve the model.

Through inspection of Figure 6, it needs to be mentioned that the [Cu/Fe] abundance ratios of Johnson et al. (2014) show apparent large scatter when [Fe/H] $> -0.6$ dex. McWilliam (2016) suspected that this scatter may be due to the measurement uncertainty or inhomogeneous chemical evolution. If the scatter is real, it may indicate that the bulge composition evolved much more inhomogeneously than that of the Galactic disk. However, this scatter is neither confirmed by Ernandes et al. (2018) or our results, which suggests that it more likely comes from measurement uncertainty.

## 6. Conclusion

We have analyzed the copper abundances for 129 bulge giants in the metallicity range $-1.14 <$ [Fe/H] $< +0.46$ dex. The samples were selected in order to investigate the [Cu/Fe] trend in the bulge, particularly at supersolar metallicities. Our copper abundances are derived from both LTE and NLTE calculations. Based on the results, we conclude that:

1. Our results confirm that the NLTE effects for Cu I lines show a clear dependence on metallicity found by Yan et al. (2015), and they gradually increase with decreasing [Fe/H]. The NLTE corrections are positive for the Cu I line, which means that the LTE results underestimate the Cu abundances.
2. The NLTE corrections of the Cu I line range from 0.0 to $\sim 0.2$ dex for our program stars, and it is evident that NLTE effects are important not only for metal-poor stars but also for metal-rich ones.
3. Our results show that the [Cu/Fe] trend of bulge stars is similar to that of the Galactic disk stars spanning the metallicity range $-1.14 <$ [Fe/H] $< 0.0$ dex, and it rises with increasing metallicity when [Fe/H] $> 0.2$ dex.
4. The [Cu/Fe] ratios increase with increasing metallicity when $\sim -1.2 <$ [Fe/H] $< \sim -0.5$ dex, which indicates that copper behaves as a secondary (metallicity-dependent) element.

To the best of our knowledge, this work is the first NLTE analysis of copper abundances for bulge stars. Our results indicate that NLTE effects should be taken into consideration for abundance analysis for Cu I lines in the bulge metal-poor and metal-rich stars. Moreover, we believe that our work provides helpful observational evidence for the improvement of existing chemical evolutionary models.

We acknowledge the data provided by the ESO Science Archive. This research was supported by the Key Research Program of the Chinese Academy of Sciences under grant No. XDPB09-02 and the National Key Basic Research Program of China under grant No. 2014CB845700, and the National Natural Science Foundation of China under grant Nos. 11833006, 11473033, and 11603037. This work is also supported by the Astronomical Big Data Joint Research Center, co-founded by the National Astronomical Observatories, Chinese Academy of Sciences, and the Alibaba Cloud.

## ORCID iDs

X. D. Xu  https://orcid.org/0000-0003-4789-0621
J. R. Shi  https://orcid.org/0000-0002-0349-7839